\documentclass[12pt]{article}
\pdfoutput=1
\usepackage{putex}
\usepackage{graphicx}
\usepackage{epstopdf}
\usepackage{enumerate}
\usepackage{amssymb,amscd,braket}
\usepackage{amsfonts}
\usepackage{cite}
\usepackage{tensor}
\usepackage{slashed}
\usepackage{feynmf}
\usepackage{hyperref}
\usepackage{url}
\usepackage{float}
\usepackage[bottom]{footmisc}
\usepackage[utf8]{inputenc}
\usepackage[margin=2.3cm]{geometry}
\usepackage{tikz}
\usepackage{amssymb}

\hypersetup{
	pdfstartview={FitH},    % fits the width of the page to the window
	pdftitle={Title},    % title
	pdfauthor={Besken, de Boer, Mathys},     % authors
	colorlinks=true,       % false: boxed links; true: colored links
	linkcolor=black,          % color of internal links (change box color with linkbordercolor)
	citecolor=blue!59!black! ,        % color of links to bibliography
	filecolor=magenta,      % color of file links
	urlcolor=blue!75!black,           % color of external links
	linktoc=all
}

\makeatletter
\g@addto@macro\bfseries{\boldmath}
\makeatother
\numberwithin{equation}{section}

\def\D{\Delta}

\newcommand {\be} {\begin {equation}}
\newcommand {\ee} {\end {equation}}

\newcommand {\bes} {\begin {equation*}}
\newcommand {\ees} {\end {equation*}}

\def\cF{{\cal F}}

\def\cO{{\cal O}}

\def\cQ{{\cal Q}}

\def\cV{{\cal V}}

\def\b{\beta}

\def\t{\tau}

\def\D{\Delta}

\def\La{\Lambda}
\def\La{\Lambda}

\def\tcV{\widetilde{\cV}}

\newcommand{\beq}{\begin{equation}}
\newcommand{\eeq}{\end{equation}}

\newcommand{\p}{\partial}

\def\be{ \begin{equation} }
\def\ee{ \end{equation} }

\def\cF{{\cal F}}

\def\b{\beta}

\def\qb{\overline{q}}

\def\tb{\overline{\tau}}

\newcommand{\bea}{\begin{eqnarray}}
\newcommand{\eea}{\end{eqnarray}}

\newcommand{\no}{\nonumber}

\newcommand\zb{\bar{z}}

\def\hb{\overline{h}}
\def\zb{\overline{z}}

\renewcommand{\title}[1]{\vbox{\center\LARGE{#1}}\vspace{5mm}}
\renewcommand{\author}[1]{\vbox{\center#1}\vspace{5mm}}
\newcommand{\address}[1]{\vbox{\center\footnotesize\em#1}}
\newcommand{\email}[1]{\vbox{\center\footnotesize\tt#1}\vspace{5mm}}

\newcommand{\<}{\langle}
\renewcommand{\>}{\rangle}

\DeclareSymbolFont{matha}{OML}{txmi}{m}{it}% txfonts
\DeclareMathSymbol{\vv}{\mathord}{matha}{118}

\begin{document}

\begin{titlepage}

\begin{center}

\hfill \\
%\vskip 1cm

\title{Virasoro conformal bootstrap with $c>1$}

\vspace{1cm}

\author{Mert Be\c sken}

\vspace{1cm}
\address{Institute for Theoretical Physics and Delta Institute for Theoretical Physics,\\ University of Amsterdam,\\
PO Box 94485, 1090 GL Amsterdam, The Netherlands 
}
\vspace{1cm}

\email{\href{mailto:m.besken@uva.nl}{m.besken@uva.nl}}

\vspace{2cm}

\textbf{Abstract}
\vspace{0.7cm}
\end{center}

We derive new constraints on the spectrum of two-dimensional conformal field theories with central charge $c>1.$ Employing the pillow representation of the four point correlator of identical scalars with dimension $\D_{\cO}$ and positivity of the coefficients of its expansion in the elliptic nome we place central charge dependent bounds on the dimension of the first excited Virasoro primary the scalar couples to, in the form $\D_1<f(c,\D_{\cO}).$ We give an analytic expression for $f(c,\D_{\cO})$ and write down transcendental equations that significantly improve the analytic bound. We numerically evaluate the stronger bounds for arbitrary fixed values of $c$ and $\D_{\cO}.$

\vfill

\end{titlepage}

\eject

\tableofcontents
\addtocontents{toc}{\protect\setcounter{tocdepth}{2}}

%	\begin{center}
%	
%	\institution{UvA}{$^{1}$Institute for Theoretical Physics, \cr
%		 University of Amsterdam, Amsterdam, The Netherlands}
%	
%	\title{Title}
%	
%		\authors{Mert Be\c sken$^{1}$}
%	
%	\abstract{Commutators are nice }
%	
%	\date{}
%	
%	\end{center}
%	\maketitle
%	\setcounter{tocdepth}{2}
%	\begingroup
%	\hypersetup{linkcolor=black}
%	\tableofcontents
%	\endgroup
%	%\tableofcontents
%	
	%%%%%%%%%%%%%%%%%%%
	
\section{Introduction}

Unitarity and crossing symmetry constrain correlation functions in conformal field theories. The constraint of crossing symmetry in two-dimensions leads to a complete classification of unitary CFTs with central charge $c<1$ known as minimal models \cite{BPZ}. Interest surged in the conformal bootstrap program in $d>2$ CFTs following the seminal paper \cite{Rattazzi:2008pe}. In this work a numerical implementation of the crossing constraint was made possible by truncating the crossing equation and evaluating it at a crossing symmetric point in cross ratio space. In four-dimensions the known form of conformal blocks \cite{Dolan:2000ut}, which are building blocks of correlators, makes this possible. In odd dimensions there does not exist explicit expressions for conformal blocks, but recursive relations for these make the study of crossing symmetry possible. This method and improvements thereof have led to remarkable success in CFTs in various dimensions, such as a precise determination of scaling dimensions in the 3d Ising CFT \cite{ElShowk:2012ht,Kos:2014bka,Kos:2016ysd}. For a review of related developments see \cite{Simmons-Duffin:2016gjk,Poland:2018epd}.

A closely related program is that of the modular bootstrap \cite{Hellerman:2009bu,Collier:2016cls,Friedan:2013cba}. In this line of research the partition function replaces the four point correlator and modular invariance replaces the crossing constraint. Thanks to the explicitly known form of the Virasoro characters, which are building blocks of the partition function on the torus, impressive upper bounds on the dimension of the first primary above the vacuum have been derived. The state of the art bound is establised by mapping the problem to the crossing constraint on the four point correlator in a one-dimensional CFT \cite{Mazac:2016qev,Hartman:2019pcd}.

The constraint of crossing symmetry for the four point correlator in two-dimensional CFT is difficult to implement for $c>1$ due to the lack of explicit expressions for the Virasoro conformal block. However, Virasoro blocks satisfy recursion relations as laid out by Zamolodchikov \cite{Zamolodchikov:1985ie,Zamor}, for a review see \cite{Perlmutter:2015iya} and appendix A of \cite{Chen:2017yze}. In this note we study the crossing equation with identical external scalars using the pillow representation of the four point correlator \cite{Maldacena:2015iua}. The pillow correlator is a function of the elliptic nome $q=e^{i\pi \tau}$ which we define precisely below. The remarkable property of the pillow correlator is that it has an expansion in $q$ with positive coefficients for identical external scalars \cite{Maldacena:2015iua}. Evaluating the pillow correlator at the crossing symmetric point $q=e^{-\pi}$ we are able to extract useful information from the crossing equation at arbitrary central charge $c>1$. For previous studies of the pillow correlator see for example \cite{Cardona:2021kyh,Das:2017cnv}.

Before going into details we report our result. In an arbitrary two-dimensional CFT with central charge $c>1,$ given a scalar Virasoro primary $\cO$ with dimension $\D_\cO,$ the lowest dimension non-vacuum Virasoro primary $\cO_1$ it couples to, which can be spinning or scalar, has its dimension $\D_1$ bounded above as,
\begin{align}
\label{intd1}
\D_1 < \D_b={1\over 6\pi \D_s}\left( 9\D_s(1+\pi \D_s)+\sqrt{3\D_s(c+3\D_s(1+\pi \D_s)(5+\pi \D_s))} \right)\, ,
\end{align}
where
\begin{align}
\D_s={48\D_{\cO}+(\pi-3)c\over 12\pi}>0\,.
\end{align}
We derive stronger bounds than \eqref{intd1} in the body of the paper, but these are numerical and their analytic expression requires solving a transcendental equation which we do not attempt in this work.

In section \ref{sec:pill} we review the pillow representation of the four point correlator. In section \ref{sec:cross} we write the crossing relations for the pillow correlator and derive our bounds. In appendix \ref{app:mm} we discuss the application of our bounds to minimal models and discuss the caveats. We end with conclusions.

\section{Pillow four point correlator}
\label{sec:pill}

We discuss the pillow representation of a four point function in two-dimensional CFT. On the plane the four point function
\begin{align}
\cF(z,\zb) = \lim_{z_4,\zb_4 \to \infty} z_4^{2h_4} z_4^{2\hb_4} \< \cO_1(0) \cO_2(z,\zb) \cO_3(1) \cO_4(z_4,\zb_4) \>
\end{align} 
decomposes in Virasoro blocks as
\begin{align}
\cF(z,\zb) = \sum_{h,\hb} C^2_{\cO\cO\cO_{h,\hb}} \cV_{h,h_i,c}(z) \cV_{\hb,\hb_i,c}(\zb)\,.
\end{align} 
The pillow variables are defined as \cite{Zamor,Maldacena:2015iua}
\begin{align}
q &= e^{i\pi\t} = {z\over 16}+\ldots \, ,\\
\t&= i {K(1-z)\over K(z)}\, ,\\
K(z)&= {1\over 2} \int_0^1 {dt \over \sqrt{t(1-t)(1-tz)}}={\pi \over 2}\,_2F_1(\tfrac{1}{2},\tfrac{1}{2};1;z)\, ,\\
z&= {\theta_2(q)^4 \over \theta_3(q)^4} \, .
\end{align}
A recursion relation \cite{Zamor,Chen:2017yze} for the Zamolodchikov $H$-function defined as
\begin{align}
\cV_{h,h_i,c}(z) &= (16q)^{h-{c-1\over 24}} z^{{c-1\over 24}-h_1-h_2} (1-z)^{{c-1\over 24}-h_2-h_3} \theta_3(q)^{{c-1\over 2}-4(h_1+h_2+h_3+h_4)} H_{h,h_i,c}(q)
\end{align} 
gives the expansion coefficients in 
\begin{align}
\label{zamh}
H_{h,h_i,c}(q)=1 + b_1 q  + b_2 q^2+ \ldots \, .
\end{align}
The pillow correlator is defined as
\begin{align}
\cF(z,\zb) &= \La(z) \La(\zb) g(q,\qb) \, ,\\
\La(z) &= \theta_3(q)^{{c\over 2}-4(h_1+h_2+h_3+h_4)} z^{{c\over 24}-h_1-h_2} (1-z)^{{c\over 24}-h_2-h_3}\, ,
\end{align}
and has an expansion in terms of pillow blocks
\begin{align}
g(q,\qb) = \sum_{h,\hb} C^2_{\cO\cO\cO_{h,\hb}}  \tcV_{h,h_i,c}(q) \tcV_{\hb,\hb_i,c}(\qb)\, .
\end{align}
The pillow block is related to the Zamolodchikov $H$-function as\footnote{$\eta(\t) = q^{1\over 12} \prod_{n=1}^{\infty} (1-q^{2n}).$}
\begin{align}
\tcV_{h,h_i,c}(q) &=\La(z)^{-1} \cV_{h,h_i,c}(z)\, , \\
\label{vtih}
&= 16^{h-{c\over 24}} q^{h-{c-1\over 24}} \eta(\t)^{-{1\over 2}} H_{h,h_i,c}(q)\, .
\end{align}
This leads to a $q$-expansion for the pillow block
\begin{align}
\tcV_{h,h_i,c}(q) = q^{h-{c\over 24}} \sum_{n=0}^{\infty} a_n(h) q^{n}\, .
\end{align}
The pillow block coefficients $a_n(h)$ really depend on all $h_i,c,h,$ but to reduce clutter we only emphasize its dependence on the exchanged operator dimension. Importantly, the pillow block expansion coefficients are non-negative when
\begin{align}
a_n(h) \geq 0 ~\Leftrightarrow~ h_1=h_4 \text{ and } h_2=h_3 \, .
\end{align}
This is due to the fact that in this circumstance these coefficients can be given the interpretation of norms on Hilbert space in pillow quantization\cite{Maldacena:2015iua}.

Plugging in the expansion for the block the pillow correlator reads
\begin{align}
\label{gpos}
g(q,\qb) = \sum_{h,\hb} \sum_{n=0}^{\infty}  \sum_{m=0}^{\infty} a_m(h) a_n(\hb) C^2_{\cO\cO\cO_{h,\hb}} q^{h+n-{c\over 24}}  \qb^{\hb+m-{c\over 24}} \, .
\end{align}
We focus on the case of identical scalars for external operators.\footnote{In the rest of the paper the labels $h_1,~h_2$ and all derivatives thereof refer to the first and second excited primaries appearing in the OPE $\cO \times \cO$. We denote the dimension of the Virasoro scalar primary $\cO$ by $h_{\cO}={\D_{\cO}\over 2}.$} In this case only even powers of $q$ appear in the expansion of the $H$-function \eqref{zamh} \cite{Chen:2017yze}. We use the Mathematica code presented in \cite{Chen:2017yze} to determine a fixed number of these coefficients and plug the resulting series into \eqref{vtih}. The series expansion of the resulting pillow block gives $a_n(h).$

\section{Crossing symmetry}
\label{sec:cross}

Crossing symmetry constrains the spectrum. For identical external scalars exchanging the positions of the first and third operator leads to
\begin{align}
\cF(z,\zb) = \cF(1-z,1-\zb)\, .
\end{align}
Under the transformation $z \to 1-z$ we write $\t \to \tilde{\t}= -1/\t,~\tb \to \tilde{\tb}= -1/\tb$ and the crossing equation reads
\begin{align}
\La(z) \La(\zb) g(q,q)=\La(1-z) \La(1-\zb) g(\tilde{q},\tilde{\qb}) \, .
\end{align}
To simplify this we note
\begin{align}
{\La(1-z)\over \La(z)} &= \left({\theta_3(\tilde{q}) \over \theta_3(q) }\right)^{{c\over 2}-16h_{\cO}} = \left(\sqrt{-i \t}\right)^{{c\over 2}-16h_{\cO}} = (-i\t)^{{c\over 4}-4\D_{\cO}},\quad \t \in \mathbb{H}^+ \, ,\\
{\La(1-\zb)\over \La(\zb)} &= \left({\theta_3(\tilde{\qb}) \over \theta_3(\qb) }\right)^{{c\over 2}-16h_{\cO}} = \left(\sqrt{i \tb}\right)^{{c\over 2}-16h_{\cO}} = (i\tb)^{{c\over 4}-4\D_{\cO}},\quad \tb \in \mathbb{H}^- \, ,
\end{align}
where we denote by $\mathbb{H}^{\pm}$ upper and lower half planes. Our domain assignment for $\t,\tb$ is the relevant one for the Euclidean regime we are interested in $\zb=z^*.$ The crossing equation is
\begin{align}
\label{gttb}
g(\t,\tb) = (\t \tb)^{{c\over 4}-4\D_{\cO}} g(-1/\t,-1/\tb)\, .
\end{align} 
For $\t={i\b/\pi},~\tb={-i\b/\pi}$ the nomes are
\begin{align}
q=e^{ i \pi \t} = e^{-\b },~~~\qb=e^{- i \pi \tb} = e^{-\b }\, ,
\end{align}
and the expansion reads
\begin{align}
g(\b) = \sum_{k=(h,\hb)} \sum_{n=0}^{\infty}  \sum_{m=0}^{\infty} a_m(h) a_n(\hb) C^2_{\cO\cO\cO_k} e^{-\b({E_k+m+n})}\, ,
\end{align}
where we defined\footnote{In the rest of the paper we use $E$ and $\D$ interchangebly. They are always related by the following relation.}
\begin{align}
\label{dele}
\D=h+\hb,~~~E=\D-{c\over 12}\, ,
\end{align}
and introduced the collective label $k=(h,\hb).$

As mentioned earlier only even powers of $q$ appear in the expansion of the $H$-function. Absorbing the overall constant $16^{E_k}$ into the OPE coefficients $C^2_{\cO\cO\cO_k}$ and setting the identity operator OPE coefficient $C^2_{\cO\cO\mathbb{I}}=1$ the first few terms look like
\begin{align}
&g(\b) =  e^{-\b E_0} + 2 a_2(0) e^{-\b (E_0+2)} + \left[a_2(0)^2+2 a_4(0)\right] e^{-\b (E_0+4)} + \ldots \no \\
& +C^2_{\cO\cO\cO_1}\left( e^{-\b E_1} + \left[ a_2(h_1)+a_2(\hb_1)\right] e^{-\b (E_1+2)} + \left[ a_2(h_1)a_2(\hb_1) + \left( a_4(h_1)+a_4(\hb_1)\right)  \right]e^{-\b (E_1+4)}+ \ldots\right) +\ldots \, .
\end{align}
The first non-trivial coefficient is simple enough to display
\begin{align}
a_2(h)= {16h^2 + 2h (256\D_{\cO}^2-32\D_{\cO}(c+3)+c(c+7)) +(c-16\D_{\cO})^2\over 32h^2+4h(c-5)+2c}\, .
\end{align}

Having expressed the crossing transformation as a modular transformation on the pillow, we follow a method similar to \cite{Hellerman:2009bu} to impose modular covariance. In terms of  $\t=ie^{p},~\tb=-ie^{p}$ \eqref{gttb} reads
\begin{align}
g(ie^{p},-ie^{p}) = e^{p\left({c\over 2}-8\D_{\cO}\right)} g(ie^{-p},-ie^{-p})\, .
\end{align} 
We differentiate this relation with respect to $p$ and set $p=0.$ In terms of $\b$
\begin{align}
\label{imgb}
I_m [g(\b)] \equiv \left[(-\b\p_\b)^m \big[\left(\pi/\b\right)^{{c\over 2}-8\D_{\cO}} g(\b)\big]-(\b\p_\b)^m g(\b)\right]_{\b=\pi}=0\, .
\end{align} 
The functionals $I_m$ for even $m$ are linearly dependent on the odd ones, so we take $m=1,3.$ Denoting the contribution of each conformal family by $I_m^{\cO_k}$
\begin{align}
\label{iokm}
I_m^{\cO_k} \equiv  I_m [e^{-\b E_k}] + \left( a_2(h_k)+a_2(\hb_k)\right)  I_m [e^{-\b (E_k+2)}] +\ldots \, ,
\end{align}
the crossing equations become
\begin{equation}
\label{fci1i3}
\begin{aligned}
I_1^{\mathbb{I}} &= - \sum_k C_{\cO\cO \cO_k}^2 I_1^{\cO_k} \, ,\\
I_3^{\mathbb{I}} &= - \sum_k C_{\cO\cO \cO_k}^2 I_3^{\cO_k}\, .
\end{aligned}
\end{equation}
Explicitly
\begin{equation}
\label{i3i1}
\begin{aligned}
I_1 [e^{-\b E}] &= {1\over 2}e^{-\pi E}(c-16\D_{\cO}+4\pi E) \, ,\\
I_3 [e^{-\b E}] &=\frac{1}{8} e^{-\pi E}\Big(
16 \pi  E+(c-16 \D_{\cO}+4 \pi
E) \left[(c-16
\D_{\cO})^2+2 \pi  E (c-16
\D_{\cO}-6)+4 \pi ^2
E^2\right]\Big)\, .
\end{aligned}
\end{equation}
Taking the ratio of the two crossing equations we get 
\begin{align}
\label{ce13}
{\sum_k C_{\cO\cO \cO_k}^2 I_3^{\cO_k} \over \sum_k C_{\cO\cO \cO_k}^2 I_1^{\cO_k}} - {I_3^{\mathbb{I}} \over I_1^{\mathbb{I}}}=0\, .
\end{align}
We continue by performing the first truncation on our crossing equation by picking just one primary $\cO_1.$ The OPE coefficients then cancel and we get
\begin{align}
\label{ce13o1}
{I_3^{\cO_1} \over I_1^{\cO_1}} - {I_3^{\mathbb{I}} \over I_1^{\mathbb{I}}}=0\, .
\end{align}
Defining the ratio $R_0(\D) \equiv {I_3[e^{-\b E}] \over I_1[e^{-\b E}]} = {I_3\left[e^{-\b \left(\D-{c\over 12}\right)}\right] \over I_1\left[e^{-\b \left(\D-{c\over 12}\right)}\right]}$ we start with exploring the truncated crossing equation
\begin{align}
\label{tc13}
R_0(\D)- R_0(0) \equiv \cQ_0(\D)=0\, ,
\end{align}
where we kept only the first term $a_0=1$ in the $q$-expansion of the Virasoro block \eqref{iokm} and defined the crossing function $\cQ_0(\D)$. $R_0(\D)$ has a pole at $\D_s$ given by
\begin{align}
\label{ds}
I_1 [e^{-\b E_s}] =0,\quad \D_s={48\D_{\cO}+(\pi-3)c\over 12\pi}>0\, ,
\end{align}
where $s$ stands for singular. Away from this pole $R_0(\D)$ is a rational convex function of the form
\begin{align}
R_0(\D) = {\pi^2 \D^3 + i_2 \D^2 + i_1 \D + i_0  \over  \D + j_0 } \, .
\end{align}
The coefficients $i,j$ depend on $c,\D_{\cO}$ and can be read from \eqref{i3i1}. Apart from $\D=0$ the truncated crossing function \eqref{tc13} has two zeros $\D_b>\D_m>0.$ The larger zero $\D_b$ has the property
\begin{align}
R_0(\D) > R_0(0),\quad \D>\D_b \, .
\end{align}
Thus, the truncated crossing function is positive for $\D>\D_b.$ The constraint implied by this on the spectrum is best expressed in terms of $\D_s$ \eqref{ds},
\begin{align}
\label{d1ds}
\D_1 <\D_b={1\over 6\pi \D_s}\left( 9\D_s(1+\pi \D_s)+\sqrt{3\D_s(c+3\D_s(1+\pi \D_s)(5+\pi \D_s))} \right) \, ,
\end{align}
where $\D_1$ is the dimension of the first excited state $\cO_1$ that is in the spectrum\footnote{In the case of minimal models identity can be the only Virasoro primary in the OPE. We comment on this in appendix \ref{app:mm}.} and couples to $\cO.$ This is a rigorous bound as we will show shortly. $\cO_1$ can be a spinning or scalar primary.

\begin{figure}[!htb]
\centering
\includegraphics[width=0.6\linewidth]{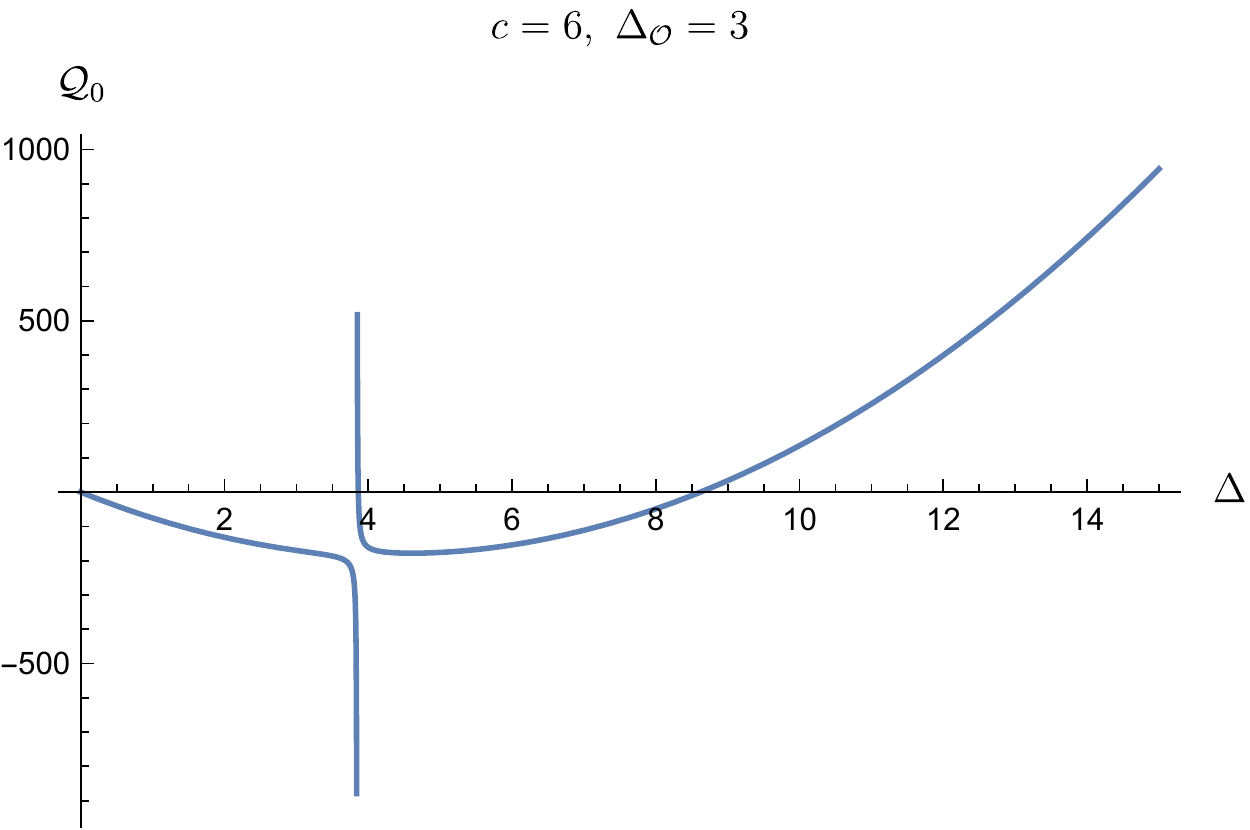}
\caption{Crossing function $\cQ_0$}
\label{fig:cfq0}
\end{figure}

To see the bound in pictures we set $c=6,~\D_{\cO}=3.$ We plot the crossing function $\cQ_0$ in Figure \ref{fig:cfq0}.
\newline

\noindent
The pole and zero of $\cQ_0(\D)$ are seen to be located at, respectively,
\begin{align}
\D_s&= 3.8423 \, ,\\
\D_b&= 8.6210 \, .
\end{align}

Next we establish the validity of the bound \eqref{d1ds} in two steps and in the process show how the bound is improved. First step is to add more terms from the $q$ expansion. Including the $q^2$ terms from \eqref{iokm} we define
\begin{align}
\label{tc13q2}
R_2(\D)- R_2(0) \equiv \cQ_2(\D)\, ,
\end{align}
where
\begin{align}
\label{r2e}
R_2(\D)={I_3 [e^{-\b E}] +\left( a_2(h)+a_2(\hb)\right) I_3 [e^{-\b (E+2)}] \over I_1 [e^{-\b E}] + \left( a_2(h)+a_2(\hb)\right) I_1 [e^{-\b (E+2)}] } \, .
\end{align}
Here we abuse notation and not emphasize the $(h,\hb)$ dependence of $R_2(\D).$ The reason for this is we can be agnostic to its $(h,\hb)$ dependence, treat $a_2(h),~a_2(\hb)$ as non-negative but otherwise arbitrary numbers and improve our previous bound $\D_b.$ To achieve this we note $R_2(\D)$ is a convex function of $\D$ away from a pole and evaluate our new crossing function at our previous bound $\cQ_2(\D_b).$ If this value is positive we will have improved our bound. Explicitly we want to show
\begin{align}
\label{q2q0c}
\cQ_2(\D_b) &> \cQ_0(\D_b)=0 \, ,\\
R_2(\D_b)- R_2(0)&>R_0(\D_b)- R_0(0)\, ,
\end{align}
or
\begin{align}
\label{r2dbi}
R_2(\D_b)-R_0(\D_b) &>R_2(0)- R_0(0)\, .
\end{align}
The formula \eqref{r2e} for $R_2(\D)$ shows explicitly how it is constructed from $R_0(\D).$ Given this construction all we need to establish \eqref{r2dbi} is the positivity of the constants $a_2(h)$ and the fact that $R_0(\D)$ is decreasing at $\D=0$ and increasing at $\D=\D_b.$ Both of these facts follow from definitions and we have established \eqref{q2q0c}. Our improved bound is located at
\begin{align}
\cQ_2(\D_b')&=0 \, ,\\
\D_1<\D_b'& \leq \D_b \, .
\end{align}
Similarly, including the $q^4$ terms, 
\begin{align}
R_4(\D)={I_3 [e^{-\b E}] + \left( a_2(h)+a_2(\hb)\right) I_3 [e^{-\b (E+2)}] + \left[ a_2(h)a_2(\hb) + \left( a_4(h)+a_4(\hb)\right)  \right] I_3 [e^{-\b (E+4)}] \over I_1 [e^{-\b E}] + \left( a_2(h)+a_2(\hb)\right) I_1 [e^{-\b (E+2)}] + \left[ a_2(h)a_2(\hb) + \left( a_4(h)+a_4(\hb)\right)  \right] I_1 [e^{-\b (E+4)}]}\, ,
\end{align}
and defining
\begin{align}
\label{tc13q4}
R_4(\D)- R_4(0) \equiv \cQ_4(\D)\, ,
\end{align}
we improve our bound
\begin{align}
\cQ_4(\D_b'')&=0 \, ,\\
\D_1<\D_b''& \leq \D_b' \, .
\end{align}

We now face the fact that $R_2,~R_4$ and as a result $\cQ_2,~\cQ_4$ are really functions of $(h,\hb).$ In practice bounds on $(h_1,\hb_1)$ come from disallowing regions in $(h,\hb)$ plane where $\cQ_n(h,\hb)$ are positive definite. These regions can be located by numerical evaluation and a three dimensional plot. Nevertheless, as our derivation has shown, no one of $h_1$ and $\hb_1$ can be too large, since their sum is strictly bounded. 

\begin{figure}[!htb]
	\centering
	\includegraphics[width=0.6\linewidth]{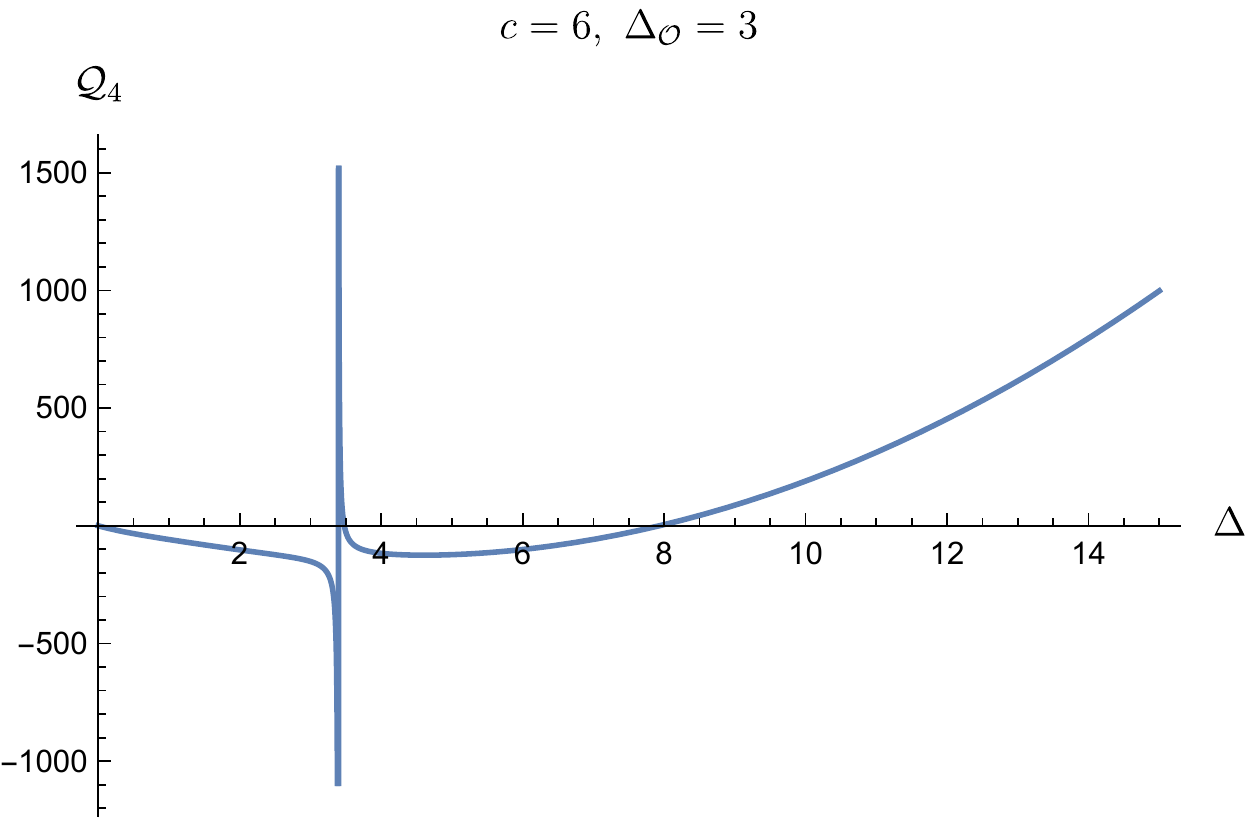}
	\caption{Crossing function $\cQ_4$}
	\label{fig:cfq4}
\end{figure}

For illustrative purposes, in our example case we assume $\cO_1$ is a scalar primary. In this case our crossing functions are really functions of $\D$ and we can give a two dimensional plot. In Figure \ref{fig:cfq4} we plot  $\cQ_4(\D)$.
With the help of Mathematica we numerically evaluate the tighter bound
\begin{align}
\label{c6d3}
\D_1&< 7.9326\, .
\end{align}
Adding more terms in the $q$ expansion of the crossing equation improves the bound \eqref{c6d3} in miniscule amounts. For example the next term  $q^6=e^{-6\pi} \approx 6.5 \times 10^{-9}$ is tiny compared to rest of the terms. 

The last step in establishing the validity of our bounds is to incorporate the rest of the primaries and show that the bounds stay intact. To demonstrate this we go back to our first crossing function \eqref{tc13} and define
\begin{align}
R_0^{(2)}(\D)- R_0(0) \equiv \cQ_0^{(2)}(\D) \, ,
\end{align}
where
\begin{align}
R_0^{(2)}(\D)={I_3 [e^{-\b E}] + c_2 I_3 [e^{-\b E_2}] \over I_1 [e^{-\b E}] + c_2 I_1 [e^{-\b E_2}] } \, .
\end{align}
Here
\begin{align}
\label{c2e2}
c_2={C_{\cO\cO \cO_2}^2\over C_{\cO\cO \cO_1}^2}>0,\quad \D_2>\D\, .
\end{align}
We need to show
\begin{align}
\cQ_0^{(2)}(\D_b) > \cQ_0(\D_b)=0 \, ,
\end{align}
but this follows trivially from the assumptions \eqref{c2e2}. Following the same steps we went through in the absence of $\cO_2$ and incorporating successive terms in the $q$-expansion in a synchronized manner for the new primary $\cO_2$ we establish the validity of our bounds.

By induction incorporation of an infinite number of primaries does not spoil our bounds, provided the infinite sum is convergent. The $q$ expansion of the correlator is convergent on the unit $q$ disk and we are well inside it $q=e^{-\pi}.$ The same is true of $I_m [g(\b)]$.

\section{Discussion}

We initiated the study of crossing symmetry for four point correlators of identical scalars in two-dimensional CFTs with finite central charge $c>1$. Even though explicit expressions for Virasoro conformal blocks are not known, we were able to use Zamolodchikov recursion relations which give an expansion of the blocks in the elliptic nome to our benefit. Evaluating the crossing equation at the crossing symmetric point $q=e^{-\pi}=0.04$ we have seen only a few terms in the $q$-expansion suffices to get powerful bounds. 

The key point in the analysis was the non-negativity of the $q$-expansion coefficients due to the matrix element interpretation of the four point correlator in pillow quantization. Given a scalar primary  $\cO$ with dimension $\D_{\cO}$ we proved for $c>1$ that the first excited Virasoro primary in the OPE $\cO \times \cO = \mathbb{I} + \cO_1+\ldots$ has to have its dimension bounded above by $\D_b$ displayed in equation \eqref{intd1}. We gave a systematic procedure to improve this bound significantly and evaluated the improved bounds numerically at fixed values of $c,\D_{\cO}.$ We have seen these bounds are close to being saturated in certain minimal models, even though favorable conditions are required to apply the procedure to minimal models and the method is rigorous really for $c>1$ theories.

One obvious way to improve our bounds is to consider higher derivatives of the crossing equation which we plan to pursue further. We plan to investigate applications of our method to large central charge CFTs in a different occasion.

\section*{Acknowledgements}

We thank Per Kraus for valuable comments on a draft of the paper. M.B. is supported by the ERC starting grant {\scriptsize{GENGEOHOL}} (grant agreement No 715656). 

\appendix	

\section{Minimal models}
\label{app:mm}

In this appendix we discuss the application of our bounds to minimal models. In fact this is illegal, because the expansion coefficients $a_n(h)$ are not positive definite for minimal model values of external dimensions and $c<1$. However in cases where they are ``sufficiently positive", that is when they are positive in a sufficiently large neighborhood of $\D_1$ we get valid bounds\footnote{When this is not the case we get no bound.}. Another issue to be wary of is the possibility of only the identity operator appearing in the OPE. For example in the Ising model $\varepsilon \times \varepsilon = \mathbb{I}$ where $\varepsilon$ is the energy density operator. In this case the right hand sides of the crossing equations \eqref{fci1i3} vanish and the division we make in subsequent discussion is invalid. Therefore below we focus on cases where at least one excited primary is in the OPE of our identical external scalars.

\begin{figure}[!htb]
	\centering
	\includegraphics[width=0.6\linewidth]{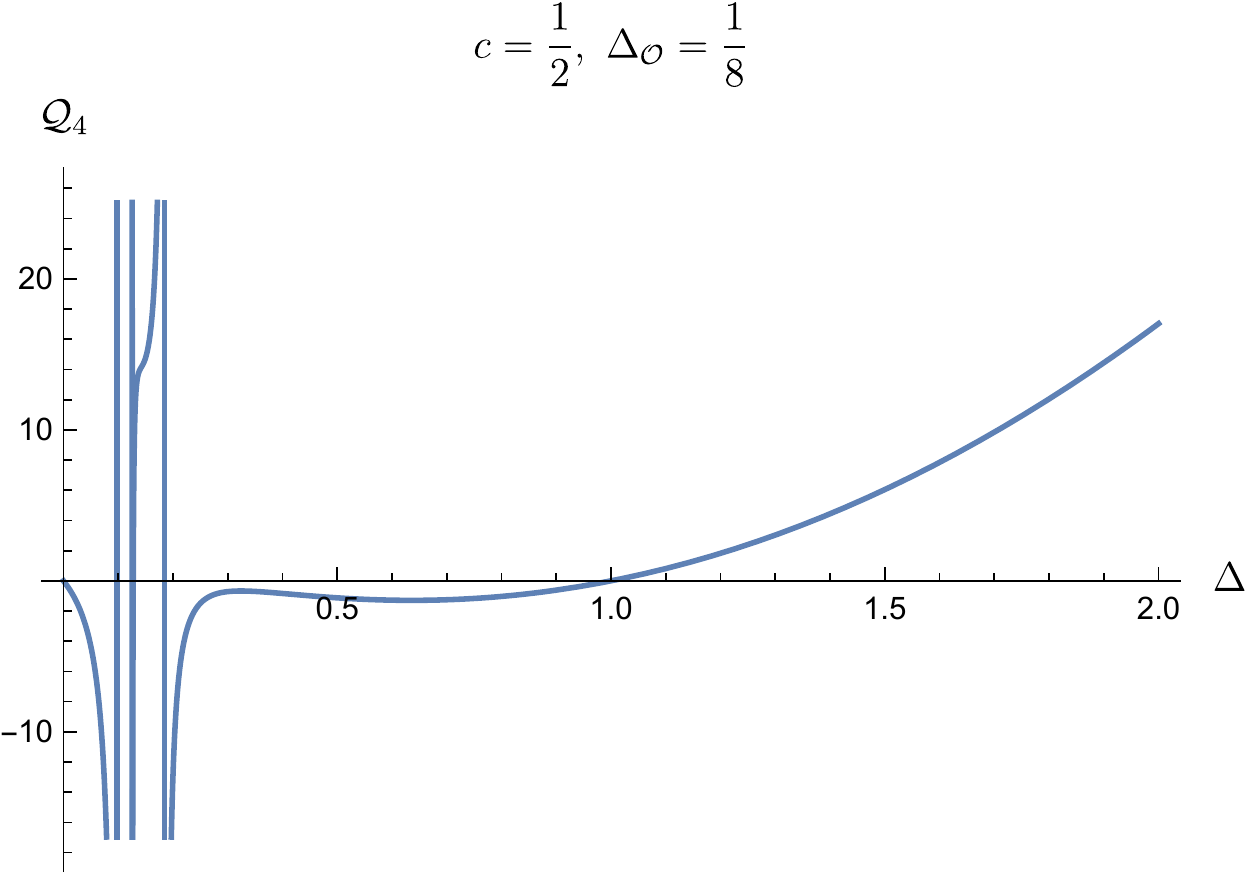}
	\caption{Crossing function $\cQ_4$ for the critical Ising model.}
	\label{fig:cfq4is}
\end{figure}

We first give the example of the 2d Ising CFT with $c={1\over 2}.$ We pick as our external scalar the spin field $\D_{\cO}={1\over 8}.$ The bound \eqref{d1ds} reads $\D_b=1.2051.$ This bound is significantly improved when we incorporate the $q^2$ and $q^4$ terms in the crossing equation. We plot $\cQ_4$ defined in \eqref{tc13q4} for the Ising CFT in Figure \ref{fig:cfq4is}.
We numerically determine the $\D$ intercept in five significant digits as
\begin{align}
\D_1<1.0001\, .
\end{align}
This is very close to the exact value of the dimension of the energy density operator $\D_1=1.$

\begin{figure}[!htb]
	\centering
	\includegraphics[width=0.6\linewidth]{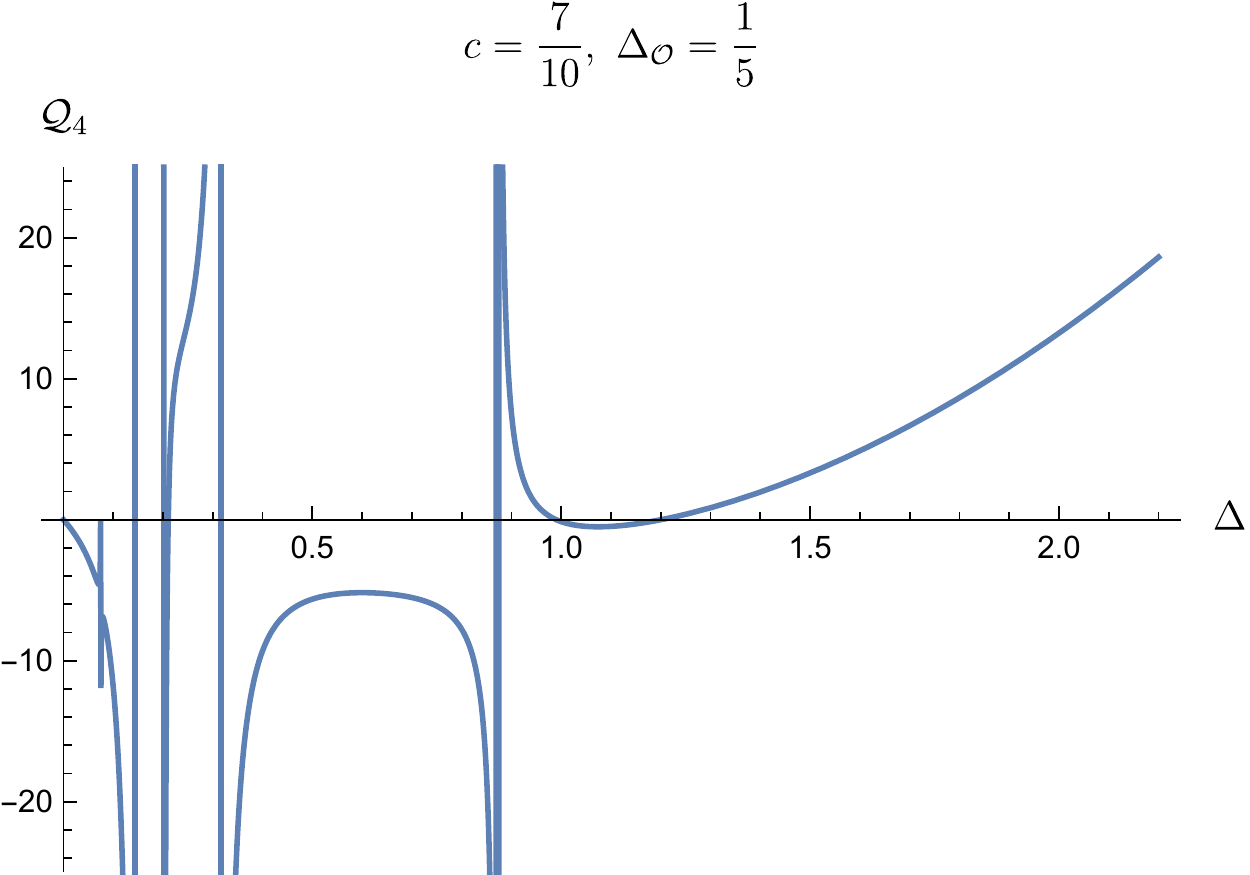}
	\caption{Crossing function $\cQ_4$ for the tricritical Ising model.}
	\label{fig:cfq4tis}
\end{figure}

Next we consider the tricritical Ising model with $c={7\over 10}.$ We pick as our external scalar the thermal operator $\D_{\cO}={1\over 5}.$ The bound \eqref{d1ds} reads $\D_b=1.4010.$ We plot $\cQ_4$ defined for the tricritical Ising model in Figure \ref{fig:cfq4tis}.
We numerically determine the $\D$ intercept in five significant digits as
\begin{align}
\D_1<1.2002\, ,
\end{align} 
which is very close to the exact value of the dimension of the thermal operator $\D_1={6\over 5}.$

%\newpage

\bibliographystyle{ytphys}
\bibliography{ref}

\end{document}